\documentclass[twocolumn,pre,amsmath,amssymb,superscriptaddress]{revtex4}

\usepackage{amsmath}    
\usepackage{amsfonts}
\usepackage{amssymb}
\usepackage{graphicx}   
\usepackage{color}
\usepackage{dcolumn}
\usepackage{bm}

\begin{document}

\title{The linear Rayleigh-Taylor instability with foams}

\author{A. Bret}
\affiliation{ETSI Industriales, Universidad de Castilla-La Mancha, 13071 Ciudad Real, Spain}
 \affiliation{Instituto de Investigaciones Energ\'{e}ticas y Aplicaciones Industriales, Campus Universitario de Ciudad Real,  13071 Ciudad Real, Spain.}
 \affiliation{MIT Plasma Science and Fusion Center, Cambridge, Massachusetts 02139, USA}
 \email{antoineclaude.bret@uclm.es.}

\author{A.J. DeVault}
\affiliation{MIT Plasma Science and Fusion Center, Cambridge, Massachusetts 02139, USA}

\author{S. G. Dannhoff}
\affiliation{MIT Plasma Science and Fusion Center, Cambridge, Massachusetts 02139, USA}

\author{M. Gatu Johnson}
\affiliation{MIT Plasma Science and Fusion Center, Cambridge, Massachusetts 02139, USA}

\author{C.K. Li}
\affiliation{MIT Plasma Science and Fusion Center, Cambridge, Massachusetts 02139, USA}

\author{J. A. Frenje}
\affiliation{MIT Plasma Science and Fusion Center, Cambridge, Massachusetts 02139, USA}

\date{\today }

\begin{abstract}
We analyse the behaviour of the linear phase of the Rayleigh-Taylor instability (RTI) in the presence of a foam. Such a problem may be relevant, for example, to some inertial confinement fusion (ICF) scenarios such as foams within the capsule or lining the inner hohlraum wall. The foam displays 3 different phases: by order of increasing stress, it is first elastic, then plastic, and then fractures. Only the elastic and plastic phases can be subject to a linear analysis of the instability. The growth rate is analytically computed in these 2 phases, in terms of the micro-structure of the foam. In the first, elastic, phase, the RTI can be stabilized for some wavelengths. In this elastic phase, a homogenous foam model overestimates the growth because it ignores the elastic nature of the foam. Although this result is derived for a simplified foam model, it is likely valid for most of them. Besides the ICF context considered here, our results could be relevant to many fields of science.
\end{abstract}

\maketitle

\section{Introduction}
With ignition reached several times at Livermore, inertial confinement fusion enters a new era where the goal is clearly to increase the yield and the repetition rate \cite{Zylstra2022,ICF2022PRL,ICF2024PRL}. In this respect, the use of foams in the target has been contemplated for some time by some authors as a means to increase laser-target coupling and to more easily and cheaply mass-produce targets compared to what is possible with solid ice layered ones \cite{Haines1997,DepierreuxPRL2009,GoncharovPRL2020,PaddockPRE2023}. Hence, in connection to ICF, foams have been studied theoretically \cite{Guskov1997,Guskov2011,Guskov2015,Belyaev2018}, numerically \cite{Kapin2006,Velechovsky2016,Belyaev2020,Milovich2021,Hudec2023} and experimentally \cite{Limpouch2006,Limpouch2008,Depierreux2009,Nicolai2012,Cipriani2018,Tikhonchuk2019,Cipriani2021,Igumenshchev2023}.

The challenge in simulating the foam behaviour lies in the various scales involved in the process. Resolving the microscopic structure of the foam during irradiation and implosion is computationally demanding \cite{FoamIFSA2025}. In this respect, the foam is often modelled as a uniform medium even though it is not, at least at the beginning of the irradiation.

In parallel, it has been recognized for long that a paramount process during the target implosion is the Rayleigh-Taylor Instability (RTI - See \cite{Hurricane2023RvMP} and references therein). In this respect, the question surges immediately: how does the RTI behave when a foam is involved? In ICF experiments, the foam is anticipated to ionize and homogenize rapidly, and so at one end of the theoretical spectrum, one can answer the question ignoring the microstructure of the foam, considering it a homogeneous medium of a designated average density. At the other end of the same spectrum, the behaviour of the RTI when an intact foam is involved, is an open question that we here address.

Experimental evidence of foam pore structure spatial imprint on transmitted shock fronts has recently been observed. In planar shock uniformity experiments conducted at OMEGA upon wetted foam samples, imaging of transmitted shock fronts with the Omega High Resolution Velocimeter revealed a spatial imprint of various pore structures upon the shock front after breakout from the back of the foam material \cite{DeVault_In_Prep}. This observation suggests that the discrete structure of foams does influence shock front progression, making clear the necessity for the development of models which describe the impact of foam pore structures on shock front progression and instability growth.

The present work does \emph{not} aim at filling the theoretical gap between intact and homogenised foam, but at exploring the ``intact’’ end of the gap. Namely, how does the RTI behaves when an intact foam is involved? To which extent does it differ from that of a homogenous fluid? Though an intact foam framework is not expected to remain
valid throughout a laser-driven ICF implosion, this work defines the limiting behavior of intact foams and explores to what extent it differs from that of a homogeneous fluid, providing the two end points of the foam-RTI spectrum.

Clearly, the foam will certainly not be intact in ICF conditions. Yet, the “imprint experiment” \cite{DeVault_In_Prep} shows that the foam may not be 100\% homogenized either. We therefore think that in order to investigate the behavior of the foam in real conditions, elucidating the other extreme state, namely the intact state, may prove useful.

Notably, a foam can be ``dry’’ or ``wetted’’. The latter would be comparable to a wetted sponge. Even though wetted foams are more relevant to ICF than dry ones \cite{PaddockPRE2023}, we shall here consider dry foams, which, for instance,
have been used to line the inner wall of hohlraums at the
NIF as a possible approach to tamping hohlraum wall
expansion and improving drive symmetry control \cite{MoorePoP2020,MoorePRE2020,Bhandarkar2018}.

To our knowledge, there is currently no theory of the mechanical properties of wetted foams, while there is for dry foams. This is why the ``wetted counterpart’’ of the present work is left for future works.

This article is structured as follows:
\begin{itemize}
  \item Presentation of  the model of foam implemented in this work, Section \ref{sec:model}.
  \item Presentation of RTI formalism implemented in this work, Section \ref{sec:forma}.
  \item Analysis of the RTI in the presence of a foam, Section \ref{sec:formafoam}.
\end{itemize}

Our main finding is as follows: as they are deformed, foams first exhibit an elastic phase with a spring-like behaviour. Then, they go through a plastic phase (see Figure \ref{fig:streerstrain}). In the elastic phase, the growth of the RTI is reduced, and even suppressed beyond a critical $k$ given by Eq. (\ref{eq:StabElas}). Analytical results are presented, for a specific model of foam, in terms of the foam inner geometry. 

The taming of the RTI in elastic materials is not new \cite{Dimonte1998PhRvL,Piriz2005,Piriz2009PRE,Piriz2009JAP}. What we think is new is that such a taming applies to foams. And likely to all of them, for all of them share the same elastic behavior for small strain \cite{Gibson1982_2D,Gibson1982_3D}.

Besides the ICF context considered here, this work could be relevant for soft matter physics \cite{YU2018,Shabalina2019}, laboratory astrophysics \cite{Kuranz2009,Rigon2021}, material science \cite{STEWART2013}, engineering \cite{Scriven1960,Grieves1963,Stewart2025}, combustion \cite{WANG2021} or geophysics \cite{Eichelberger1980,Houseman1997}.

\begin{figure}
\begin{center}
 \includegraphics[width=0.45\textwidth]{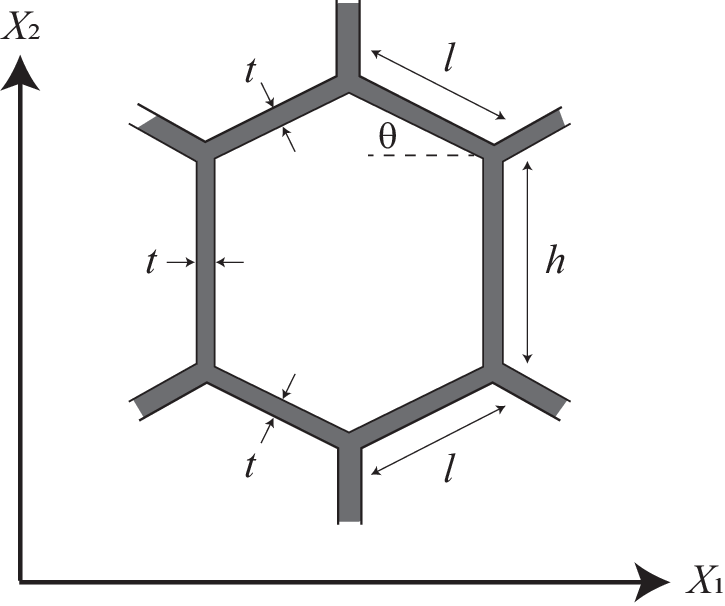}
\end{center}
\caption{Model of cell of a foam in 2D. Beams of a material of density $\rho_s$ connected to each other according to the displayed geometry. The whole structure is obtained replicating this unit in every direction. From \cite{Gibson1982_2D}.}\label{fig:foam}
\end{figure}

\section{Model of foam}\label{sec:model}
Foams come with a great variety of flavour, like two-dimensional honeycomb or three-dimensional foams, with open or closed cells, etc. We shall here focus on a two-dimensional honeycomb, the cell unit of which is represented in Figure \ref{fig:foam}.

The ``single most important feature''\footnote{In the words of Ref. \cite{Gibson1997}, page 2.} of a foam is its relative density. With the notations defined on Figure \ref{fig:foam}, it reads \cite{Gibson1982_2D},
\begin{equation}\label{eq:relat_dens}
  \frac{\rho}{\rho_s} = \frac{(2+h/l)t/l}{2 \cos \theta (h/l + \sin \theta)}.
\end{equation}
This is the average density $\rho$ of the foam, divided by the density $\rho_s$ of the material it is made of. When a foam is assimilated to an homogenous medium, the density of the equivalent homogenous medium is the density $\rho$. The impact of $\rho_s, \theta, l$ or $h$ on any process, is therefore lost.

For a regular pattern with $\theta=30^\circ$ and $h=l$, the foam stress tensor is isotropic (see below). In such a case,  its relative density reduces to
\begin{equation}\label{eq:relat_dens_reg}
  \frac{\rho}{\rho_s} = \frac{2}{\sqrt{3}}\frac{t}{l}.
\end{equation}
Since in general $t \ll l$, $\rho/\rho_s \ll 1$.

\begin{figure}
\begin{center}
 \includegraphics[width=0.45\textwidth]{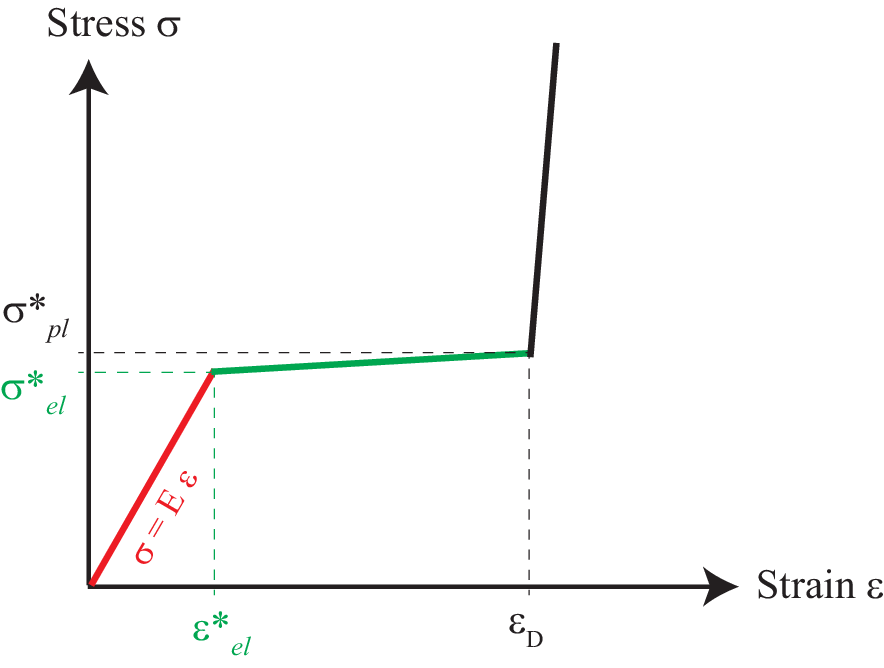}
\end{center}
\caption{Typical stress-strain curve of a foam. See Eq. (\ref{eq:strain_el_iso}) for $\varepsilon^*_{el}$ and Eq. (\ref{eq:EpsDens}) for $\varepsilon_D $. The value of $E$, slope of the red line, is given by Eq. (\ref{eq:YoungIso}). Adapted from \cite{Gibson1982_2D}.}\label{fig:streerstrain}
\end{figure}

As we shall explore the foam behaviour under the RTI, we need to know about the foam mechanical properties. They are well illustrated by the stress-strain curve of Figure  \ref{fig:streerstrain}. The stress $\sigma$ has the units of a pressure. For a material of length $L$ compressed by a length $\delta$, the strain is defined by
\begin{equation}\label{eq:straindef}
 \varepsilon \equiv \frac{\delta}{L}.
\end{equation}
 The curve shows 3 distinct stages:
\begin{itemize}
  \item The \emph{elastic} phase. For small strain, the foam acts like a spring, with a Hooke's law $\sigma = E \varepsilon$, where $E$ is the Young's modulus. Without any further assumption, directions $X_1$ and $X_2$ may have different Young's modulus.
  \item The \emph{plastic} phase. The inner structure starts to collapse. This is the quasi-plateau phase when the stress remains nearly constant as the strain keeps increasing.
  \item The \emph{fracture}  phase. The inner structure collapsed, like opposite inner walls touching each other.
\end{itemize}
We shall now review the properties of each phase.

\subsection{Elastic phase}
For a stress applied in the $X_1$ direction, the Young modulus reads (\cite{Gibson1997}, p. 102),
\begin{equation}\label{eq:YoungX1}
  E^*_1 = E_s \left( \frac{t}{l}\right)^3\frac{\cos\theta}{(h/l + \sin \theta)\sin^2\theta},
\end{equation}
where $E_s$ is the Young modulus of a beam.

For a stress applied in the $X_2$ direction, the Young modulus reads (\cite{Gibson1997}, p. 103),
\begin{equation}\label{eq:YoungX2}
  E^*_2 = E_s \left( \frac{t}{l}\right)^3 \frac{h/l + \sin \theta}{\cos^3\theta}.
\end{equation}

For a regular pattern with
\begin{eqnarray}\label{eq:foamparam}
  \theta &=& 30^\circ, \\
  h &=& l, \nonumber
\end{eqnarray}
the stress tensor is isotropic.  $E^*_1$ and $E^*_2$ then reduce to,
\begin{eqnarray}\label{eq:YoungIso}
  E^*_1 &=& E^*_2  \nonumber \\
  &\equiv & E = E_s \left( \frac{t}{l}\right)^3 \frac{4}{\sqrt{3}},
\end{eqnarray}
which is therefore the slope of the red line on Figure \ref{fig:streerstrain}.

\subsection{Plastic phase}
The plastic phase arises from the buckling of the cells walls, allowing further strain at almost constant stress. In the $X_2$ direction it occurs for the critical stress (\cite{Gibson1997}, p. 106),
\begin{equation}\label{eq:sigma_el}
   \sigma^*_{el} = E_s\frac{n^2 \pi^2}{24}\frac{t^3}{lh^2}\frac{1}{\cos\theta},
\end{equation}
where $n \in [0.5,2]$ is the so-called  ``end constraint factor'', a function of the internal foam structure. For a regular pattern with $\theta=30^\circ$ and $h=l$, $n=0.69$ and Eq. (\ref{eq:sigma_el}) reduces to \cite{Gibson1982_2D},
\begin{equation}\label{eq:sigma_el_iso}
    \sigma^*_{el} = E_s \left( \frac{t}{l} \right)^3 \frac{(0.343 \pi)^2}{3\sqrt{3}}.
\end{equation}

Putting together Eqs. (\ref{eq:YoungIso},\ref{eq:sigma_el_iso}), we can derive the elastic collapse strain $\varepsilon^*_{el}$ corresponding to such a stress,
\begin{eqnarray}\label{eq:strain_el_iso}
 E_s \left( \frac{t}{l}\right)^3 \frac{4}{\sqrt{3}}  \varepsilon^*_{el} & \equiv & E_s \left( \frac{t}{l} \right)^3 \frac{(0.343 \pi)^2}{3\sqrt{3}} \nonumber\\
  \Rightarrow  \varepsilon^*_{el} & = & \frac{(0.343 \pi)^2}{12} \sim \frac{1}{10.4}.
\end{eqnarray}
Such a low value of the maximum strain in this phase is relevant to the forthcoming instability analysis. It implies that the strain remains small all along the elastic phase, so that the linear approximation definitely applies  for this 2D hexagonal foam.

\subsection{Fracture phase}
Plastic collapse occurs at a critical stress $\sigma^*_{pl}$ where the internal structure simply collapses, and opposite walls touch each other. From this point, further increase of the stress yields no further compression, hence the nearly vertical line in Figure \ref{fig:streerstrain}.

For a regular pattern with $\theta=30^\circ$ and $h=l$,  $\sigma^*_{pl}$ reads \cite{Gibson1982_2D}
\begin{equation}\label{eq:sig_coll}
\sigma^*_{pl} = \frac{2}{3}\left( \frac{t}{l} \right)^2\sigma_y,
\end{equation}
where $\sigma_y,$ is the yield stress of the cell-wall material.

We shall model the green plateau by a horizontal line, implying, from Eqs. (\ref{eq:sigma_el_iso}, \ref{eq:sig_coll})
\begin{eqnarray}\label{eq:innerratio}
   \sigma^*_{el} &=& \sigma^*_{pl}  \nonumber\\
    \Rightarrow E_s \left( \frac{t}{l} \right)^3 \frac{(0.343 \pi)^2}{3\sqrt{3}} &=& \frac{2}{3}\left( \frac{t}{l} \right)^2\sigma_y \nonumber\\
        \Rightarrow \frac{t}{l}   &=&     \frac{2\sqrt{3}}{(0.343 \pi)^2}\frac{\sigma_y}{E_s} \sim 3 \frac{\sigma_y}{E_s}.
\end{eqnarray}
The  green plateau reaches an end at the ``densification strain'' $\varepsilon_D$ given by (\cite{Gibson1997}, p. 131)
\begin{equation}\label{eq:EpsDens}
\varepsilon_D = 1 - 1.4 \frac{(2+h/l)t/l}{2\cos\theta(h/l + \sin\theta)}.
\end{equation}

For a regular pattern with $\theta=30^\circ$ and $h=l$,  $\varepsilon_D$ reads
\begin{equation}\label{eq:EpsDensISo}
\varepsilon_D = 1 - 1.4 \binom{2}{\sqrt{3}} \frac{t}{l} \sim 1 - 1.61 \frac{t}{l}.
\end{equation}
With $t \ll l$, we obviously have $\varepsilon^*_{el} < \varepsilon_D$, where $\varepsilon^*_{el}$ is defined by Eq. (\ref{eq:strain_el_iso}). We thus check that on Figure \ref{fig:streerstrain}, $\varepsilon^*_{el}$ and $\varepsilon_D $ are correctly ordered.

Such a high value of the maximum strain in this phase is equally relevant to the forthcoming instability analysis. It implies that by the end of the plateau, the strain is necessarily close to unity, rendering the linear approximation invalid.

In summary, we here focus on foams with the following properties:
\begin{itemize}
  \item Intact foam, that is, not pre-deformed nor partially or fully homogenized by anything (laser, ablator pressure,\ldots).
  \item Inner aspect ratio fixed by Eq. (\ref{eq:innerratio}).
  \item Dry foam.
  \item 2D foam, isotropic with $\theta=30^\circ$ and $h=l$.
\end{itemize}

\begin{figure}
\begin{center}
 \includegraphics[width=0.45\textwidth]{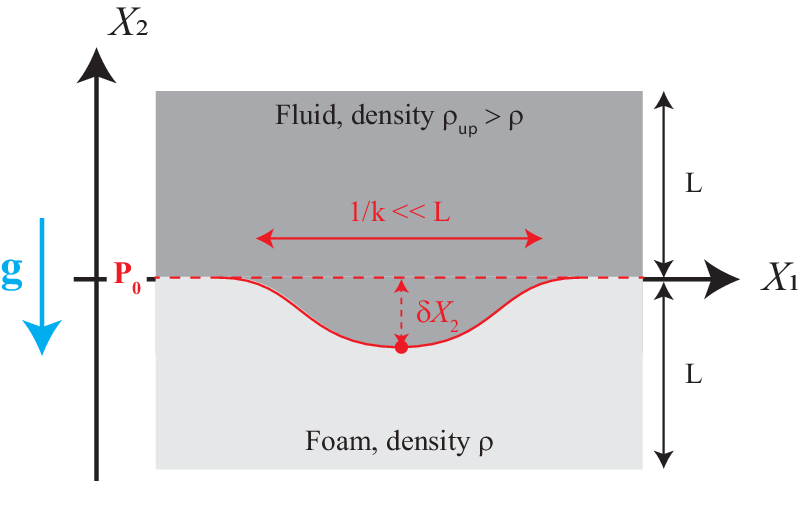}
\end{center}
\caption{Setup considered for the RTI. The foam average density is $\rho$. It is placed below a fluid of density $\rho_{up} > \rho$.}\label{fig:setup}
\end{figure}

\section{RTI formalism}\label{sec:forma}
The analysis of the RTI through the usual ``normal modes'' method can be found in various treatises \cite{chandra1981,faber1995,TB2017}. It consists in writing the fluid equations on each sides of the interface, linearizing them for small perturbations of the interface and applying some continuity requirements at the interface.

The RTI analysis we are about to present is ``non-standard’’, so to speak. It was presented in Ref. \cite{Piriz2005}, and we briefly reproduce it here.
Its advantage over the usual normal modes formalism is double: it is much more flexible and above all, much more intuitive.

Notably, it has already been applied to elastic-plastic media, with the outcome successfully tested through numerical simulations \cite{Piriz2009JAP,Piriz2009PRE} or against a rigorous theoretical approach \cite{Terrones2005}. It has even been used to retrieve the normal modes result for the relativistic RTI \cite{Bret2011LPB,Allen1984MNRAS}.

Note that while foams are not continuous media, we just saw that they behave like elastic-plastic ones. RTI works on such substances are therefore relevant to our purpose.

We shall now explain this ``non-standard’’ formalism, before applying it to the foam case.

Consider the setup pictured in Figure \ref{fig:setup}. The foam average density is $\rho$. It is placed below a fluid of density $\rho_{up} > \rho$. At equilibrium, the pressure at the interface of the two media is $P_0$. The interface is now bent over a distance $\sim 1/k \ll L$, by an amplitude $\delta X_2$. We assume the foam uniformly supports the higher density fluid above it, so that the pore size $\sim 2l\cos\theta = l \sqrt{3}$ (see Figure \ref{fig:foam} for $\theta=30^\circ$) is smaller than the wavelength of the perturbation, namely
\begin{equation}\label{eq:condk}
1/k \gg l \sqrt{3} .
\end{equation}

What is now the pressure above and below the red spot located at the lowest point of the perturbation?
\begin{itemize}
  \item The pressure \emph{above} is now $P_{ab}=P_0+\rho_{up}~g\delta X_2$.
  \item The pressure \emph{below} is now $P_{be}=P_0+\rho ~ g\delta X_2$.
\end{itemize}
Because $\rho_{up} > \rho$, it is obvious that $P_{ab} > P_{be}$: the perturbation is amplified. On the contrary, we would have $P_{ab} < P_{be}$, pushing the interface back up, and restoring its initial position.

Let us now compute the ``classical'' (no foam) linear growth rate from this simple picture. To this extent, suppose the interface has extension $D$ in the transverse, $X_3$ direction. The surface of the perturbation is therefore $S \sim D/k$. The force acting upon it reads,
\begin{eqnarray}\label{eq:FRIT}
F &=& (P_{ab} -  P_{be})S  \nonumber \\
&=& (\rho_{up}-\rho)~g~\delta X_2 S \nonumber \\
&=& (\rho_{up}-\rho)~g~\delta X_2 \frac{D}{k}.
\end{eqnarray}
oriented downward along $X_2$.

We shall now assess the total mass $M$ involved in the process and apply Newton's law. What is the total mass displaced? On both sides of the interface, it is proportional to $S$ and to the height of the layer moved, namely $1/k$ (see comments before Eq. (\ref{eq:vareps}) below). It therefore reads,
\begin{equation}\label{eq:RTIMass}
M = \rho_{up} \frac{S}{k} + \rho \frac{S}{k} = (\rho_{up}+\rho)\frac{D}{k^2}.
\end{equation}
Applying Newton's law $Ma = F$   yields,
\begin{equation}\label{eq:Newton}
(\rho_{up}+\rho)\frac{D}{k^2} \ddot{\delta X_2} = (\rho_{up}-\rho)~g~\delta X_2 \frac{D}{k},
\end{equation}
that is,
\begin{equation}\label{eq:RTIClass}
 \ddot{\delta X_2} = \gamma^2 ~\delta X_2,
\end{equation}
where,
\begin{equation}\label{eq:RTIClassGR}
\gamma^2 = \frac{\rho_{up}-\rho}{\rho_{up}+\rho} k g,
\end{equation}
which is exactly the growth rate of the ``classical'' RTI. Since Eq. (\ref{eq:RTIClass}) has solutions which are linear combinations of $\cosh (\gamma t)$ and $\sinh(\gamma t)$, the displacement $\delta X_2$ grows exponentially with time, at rate $\gamma$.

 The structure of the Atwood number,
\begin{equation}\label{eq:Atwood}
 A \equiv \frac{\rho_{up}-\rho}{\rho_{up}+\rho},
\end{equation}
 is clearly revealed: the density difference pertains to the pressure difference at the interface, and the density sum to the total mass involved in the process.

We shall now modify this treatment to account for the properties of the foam.

\section{RTI with a foam}\label{sec:formafoam}
We consider a scenario where a perturbation grows from infinitesimal amplitude  $\delta X_2=0^+$. Others are possible, like seeding it at a finite amplitude from $t=0$.

The foam properties simply modify the expression (\ref{eq:FRIT}) of the force acting upon the interface. While Eq. (\ref{eq:FRIT}) only accounts for the pressure force of each fluid, its foam counterpart needs to account, in addition, for the foam stress.

A key quantity is the foam strain, namely, the displacement of the foam interface divided by its length. Which ``length'' should be considered in this respect? It is known that in the vertical, $X_2$ direction, an interface perturbation of wavelength $k$ decays like $e^{-kX_2}$ \cite{chandra1981}. We shall then consider $1/k$ as the vertical foam length involved in the instability process, defining the strain, from Eq. (\ref{eq:straindef}), as
\begin{equation}\label{eq:vareps}
\varepsilon \equiv k \times \delta X_2.
\end{equation}
Figure \ref{fig:streerstrain} shows how the stress depends on the strain.   Since the process starts from $\delta X_2=0^+$, we shall first encounter the elastic nature of the foam, where it has a spring-like reaction to the strain.

\subsection{Limits of the linear analysis}\label{sec:limits}
Are the three phases of the foam behavior amenable to  a \emph{linear} RTI analysis?

Unstable RTI modes grow exponentially until their amplitude reaches an $\alpha$ fraction of their wavelength $\lambda = 2\pi/k$ \cite{Haan1989,Zhou2017a,Zhou2017b}, with $\alpha \in [1/10,1/5]$. Therefore, the linear regime holds until $\delta X_2 = \alpha \lambda$, that is,
\begin{equation}\label{eq:linearcrit}
k \times \delta X_2  = \varepsilon = 2\pi\alpha \in [0.6,1.2].
\end{equation}
 Figure \ref{fig:streerstrain}, with $\varepsilon^*_{el} \sim 0.1$ defined by Eq. (\ref{eq:strain_el_iso}), shows that the full elastic phase of the foam fits into the linear regime, even with $\alpha=1/10$. Part of the green plateau phase should equally fulfills the linear requirement (\ref{eq:linearcrit}).

Regarding the last phase, the fracture phase of the foam, evidenced by a nearly vertical line on Figure \ref{fig:streerstrain}, Eq. (\ref{eq:EpsDensISo}) shows it starts from $\varepsilon_D = 1 - 1.61 t/l \sim 1^-$.

Therefore, whether the full plateau and the fracture phases fit the linear regime, depends on $\alpha$. Future numerical simulation could help narrowing down its value for the present scenario.

We now assess the foam influence when the linear analysis can apply.

\subsection{Elastic phase}\label{sec:elastic}

 The force resulting from the foam stress, oriented upward along $X_2$ is the stress $\sigma$ corresponding to $\varepsilon$, times the surface $S$. According to Figure \ref{fig:streerstrain}, $\sigma = E \varepsilon$, where $E$ is the Young modulus presented in Eq. (\ref{eq:YoungIso}).  The elastic foam version of Eq. (\ref{eq:FRIT}) is therefore
\begin{eqnarray}\label{eq:FRIT_Foam}
F &=& (P_{ab} -  P_{be})S - \sigma S \nonumber \\
&=& (\rho_{up}-\rho)~g~\delta X_2 S -E k ~\delta X_2S \nonumber \\
&=& (\rho_{up}-\rho)~g~\delta X_2 \frac{D}{k} - E k~ \delta X_2\frac{D}{k} \nonumber \\
&=& \left(g - \frac{k E}{\rho_{up}-\rho} \right)(\rho_{up}-\rho)\delta X_2\frac{D}{k}.
\end{eqnarray}
Comparing with Eq. (\ref{eq:FRIT}), it appears that in the elastic phase, the effect of the foam is simply to substitute,
\begin{equation}\label{eq:subsg}
g \rightarrow g - \frac{k E}{\rho_{up}-\rho},
\end{equation}
with a growth rate of the RTI on the elastic phase,
\begin{equation}\label{eq:GR_Elastic}
\gamma '^2 = A k g \left(1 - \frac{k E}{g(\rho_{up}-\rho)}  \right),
\end{equation}
where $A$ is the Atwood number defined by Eq. (\ref{eq:Atwood}). The interface is stable against the RTI for,
\begin{equation}\label{eq:StabElas}
k > k_m \equiv g \frac{\rho_{up}-\rho}{E}.
\end{equation}
The maximum growth rate is reached for,
\begin{equation}
k =  g\frac{\rho_{up}-\rho}{2 E},
\end{equation}
with growth rate
\begin{equation}\label{eq:maxGR}
\gamma'^2 = g^2 A \frac{\rho_{up}-\rho}{4 E}.
\end{equation}
The interface stills grows exponentially for $k < k_m$, though at a lesser rate. Such a large $k$ stabilization of the RTI for elastic materials was already found in Ref. \cite{Piriz2005}. A possible physical connection to a similar growth rate reduction in an ablatively accelerating plasma could be explored \cite{Takabe1985}.

\subsection{Plastic phase}\label{sec:plastic}
In case the growth rate $\gamma$ defined by Eq. (\ref{eq:RTIClassGR}) remains positive when re-scaling $g$ according to Eq. (\ref{eq:subsg}), the perturbation will grow, with a strain $k \times \delta X_2$ reaching the green plateau on Figure \ref{fig:streerstrain}. Equally relevant to this section would be the case of a seeded perturbation with an appropriate amplitude, namely, high enough for $\varepsilon = k \times \delta X_2$ to lie on the plateau, but not too high for the linear approximation to be valid ($\varepsilon_{t=0} = 2\times 10^{-1}$ or $3\times 10^{-1}$, for example).

From this on, and until the strain leaves the linear regime, the stress  is nearly constant, equal to $\sigma_{el}^*$ defined by Eq. (\ref{eq:sigma_el}).

While the linear theory is still valid, the plastic foam version of Eq. (\ref{eq:FRIT}) is now
\begin{eqnarray}
F &=& (P_{ab} -  P_{be})S - \sigma_{el}^* S \nonumber \\
&=& (\rho_{up}-\rho)~g~\delta X_2 S - \sigma_{el}^*S \nonumber \\
&=& (\rho_{up}-\rho)~g~\delta X_2 \frac{D}{k} - \sigma_{el}^* \frac{D}{k},
\end{eqnarray}
yielding a modified equation of motion (\ref{eq:Newton}),
\begin{equation}
(\rho_{up}+\rho)\frac{D}{k^2} \ddot{\delta X_2} = (\rho_{up}-\rho)~g~\delta X_2 \frac{D}{k} - \sigma_{el}^* \frac{D}{k},
\end{equation}
that is,
\begin{equation}\label{eq:Newton_Plast}
\ddot{\delta X_2} = \gamma^2 \delta X_2  - k \frac{\sigma_{el}^*}{\rho_{up}+\rho },
\end{equation}
 with $\gamma$ still given by Eq. (\ref{eq:RTIClassGR}). This equation has solutions which, again, are linear combinations of $\cosh (\gamma t)$ and $\sinh(\gamma t)$. Hence, after the elastic phase where the growth, if happening, was slower than that of a fluid, the growth rate resumes at the fluid pace.

\begin{figure}
\begin{center}
 \includegraphics[width=0.45\textwidth]{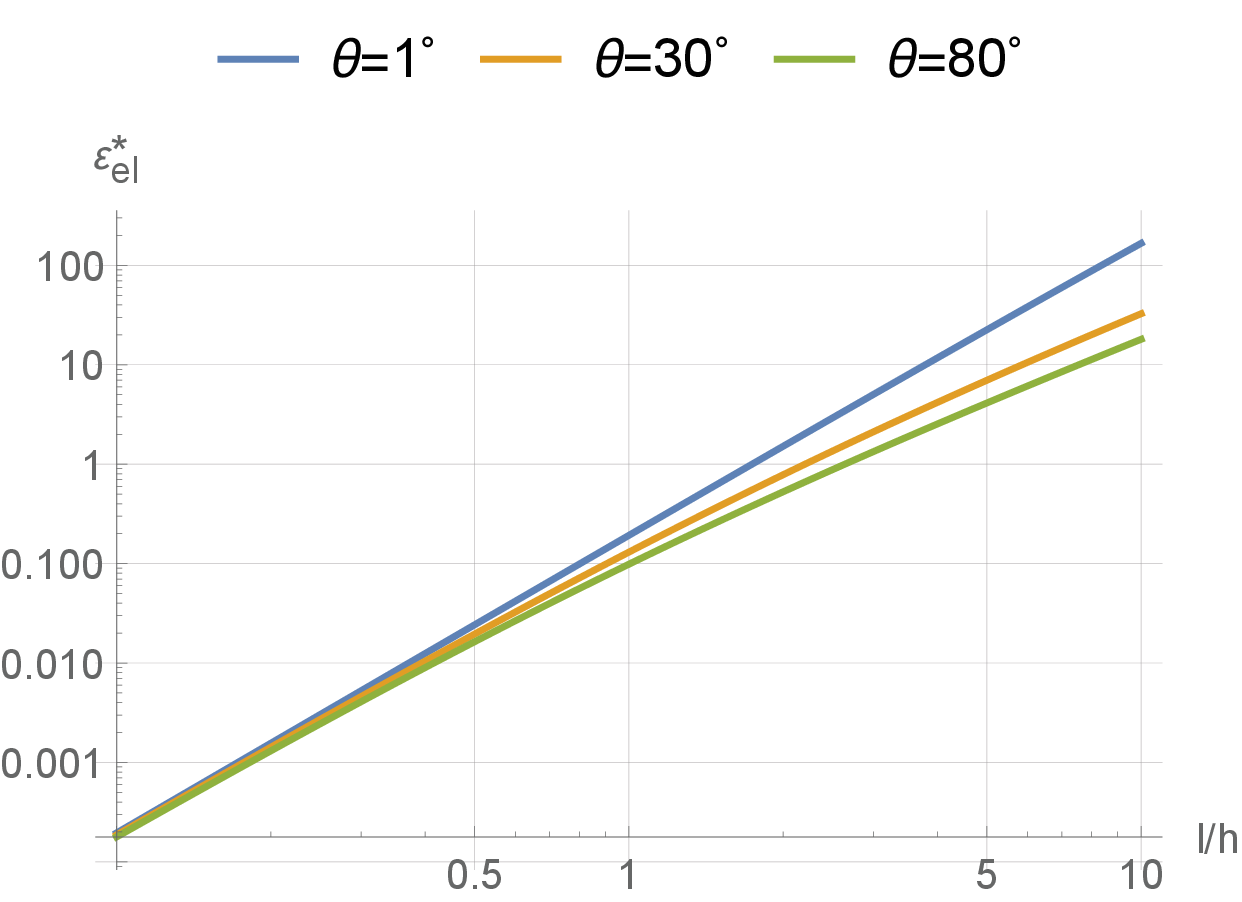}
\end{center}
\caption{Plot of Eq. (\ref{eq:general}) for 3 values of $\theta$ for $n=0.69$.}\label{fig:general}
\end{figure}

\subsection{Foam design for RTI mitigation}\label{sec:opti}
We here comment on the possibility of tailoring the foam design for applications where control of instability is important.

Since the taming of the RTI is found in the foam elastic phase, it could be beneficial for stability purposes to extend it as far as possible in terms of the strain $\varepsilon$. For the model of foam chosen here, namely $\theta=30^\circ$,  $h=l$ and $n=0.69$, $\varepsilon^*_{el}$ is set to $\sim 0.1$, as fixed by Eq. (\ref{eq:strain_el_iso}). However, coming back to the general expressions (\ref{eq:YoungX2},\ref{eq:sigma_el}) of the slope of the red line on Figure \ref{fig:streerstrain}, and of the critical stress $\sigma^*_{el}$ marking the onset of the plateau, we can obtain a more general expression of $\varepsilon^*_{el}$,
\begin{equation}\label{eq:general}
\varepsilon^*_{el} = \frac{\pi ^2 n^2}{24} \frac{(l/h)^3 \cos ^2(\theta )}{(l/h) \sin (\theta )+1}.
\end{equation}
Figure \ref{fig:general} plots this equation for $n=0.69$ and 3 values of $\theta$. The higher $l/h$, the longer the elastic phase lasts. Simply put, cells foams should be wider than high.

Varying $\varepsilon^*_{el}$ by controlling $l/h$ could help test the theory by measuring growth rates, seeing if, as predicted, an abrupt change occurs at the elastic-plastic transition.

The maximum growth rate (\ref{eq:maxGR}) in the elastic phase can can also be tailored. It reads
\begin{equation}\label{eq:GRtailor}
\gamma'^2 = g^2 A \frac{\rho_{up}-\rho}{4 E} = g^2 \frac{(\rho_{up}-\rho)^2}{\rho_{up}+\rho} \frac{1}{4 E}.
\end{equation}
Here, the inner foam design comes into play through Eqs. (\ref{eq:relat_dens},\ref{eq:YoungX2}), defining respectively the average density $\rho$ and the Young modulus $E$ (now in the $X_2$ direction). Contrary to Eq. (\ref{eq:general}), this expression involves too many parameters to be simply analyzed. However, for large values of $l/h$ (and $\theta\neq 0$), $\varepsilon^*_{el}$ given by (\ref{eq:general}) scales like $(l/h)^2$. This allows to replace $l/h$ by its expression in terms of $\varepsilon^*_{el}$ in the equation (\ref{eq:GRtailor}) above. The leading term of the result for large $l/h$, reads,
\begin{equation}
\gamma'^2 \sim g^2 \frac{12 \sqrt{6}}{\pi ^3 n^3} \frac{\rho_{up}}{E_s} \left( \frac{h}{t} \right)^3 \sqrt{\sin \theta} ~ \varepsilon^{*3/2}_{el}.
\end{equation}
It seems then that a large $\varepsilon^{*}_{el}$, that is, a long elastic phase, also favors a higher growth rate. A detailed analysis including all the factors involved is required to assess the trade off between both quantities. Note also that for foam parameters different than those defined by Eq. (\ref{eq:foamparam}),  the Young moduli in the $X_1$ and $X_2$ will be different. The anisotropy so introduced should modify the RTI analysis of Sections \ref{sec:forma} \& \ref{sec:formafoam}.

\begin{figure}
\begin{center}
 \includegraphics[width=0.45\textwidth]{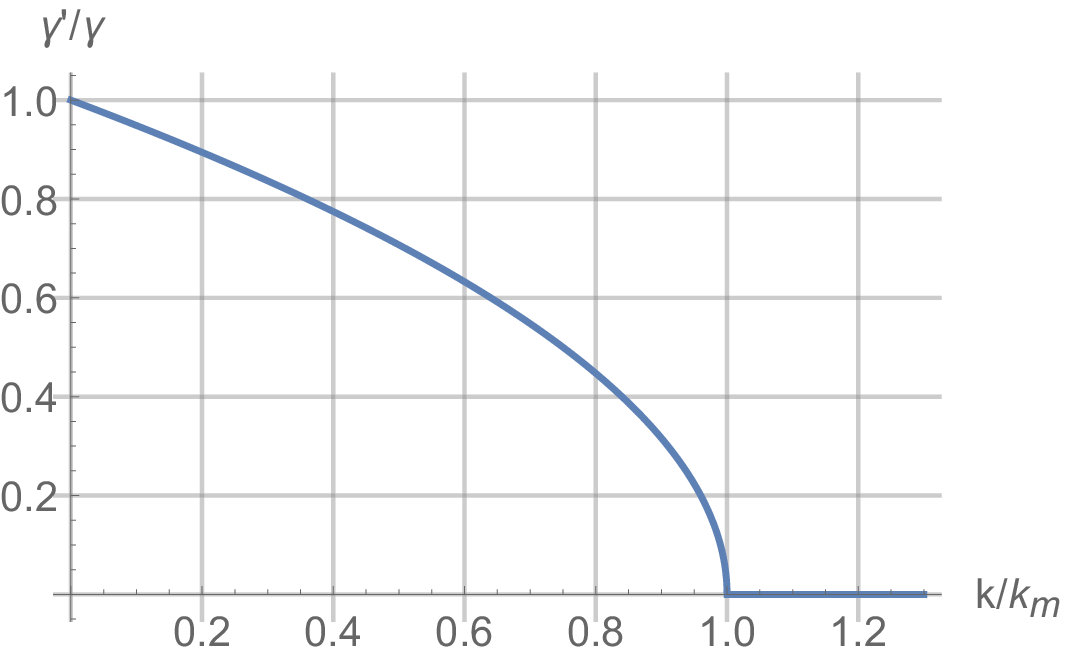}
\end{center}
\caption{Ratio of the growth rate $\gamma '$ of the foam-RTI, to the growth rate $\gamma$ of the averaged medium-RTI in the elastic phase of the foam. From Eq. (\ref{eq:gammaprime}).}\label{fig:gammaprime}
\end{figure}

\section{Conclusion}
After presenting a mechanical model of a simple foam and an intuitive description of the RTI, we came to an analytical theory of the RTI for the foam considered. As previously found, the linear phase of the RTI is relevant to 2 of the 3 mechanical phases of the foam: the elastic and the plastic phase. The last phase, the fracture one, necessarily implies too large a deformation for the linear theory to apply.

We can now assess the difference between the RTI with the foam and with the equivalent medium of average density $\rho$. Such a difference is only notable in the first phase, namely the elastic phase, since the growth rate of the RTI in the next phase, the plastic one, is the same in both cases (see Section \ref{sec:plastic}).

From Eqs. (\ref{eq:RTIClassGR},\ref{eq:GR_Elastic}), we can express the ratio of the growth rate $\gamma '$ of the foam-RTI, to the growth rate $\gamma$ of the averaged medium-RTI,
\begin{equation}\label{eq:gammaprime}
\frac{\gamma '}{\gamma} = \sqrt{1 - \frac{k}{k_m} }
\end{equation}
where $k_m$ is defined by Eq. (\ref{eq:StabElas}). This function is represented on Figure \ref{fig:gammaprime}. There is virtually no difference as small $k$'s (remember the smallest relevant $k$ is indeed $k=1/L$). For $k \lesssim k_m$, and obviously for $k > k_m$, the homogenous foam model clearly overestimates the growth, when it does not find a growth where there is not ($k > k_m$). All differences come from ignoring the elastic nature of the foam.

After replacing $\rho$ and $E$ in the expression (\ref{eq:StabElas}) of $k_m$ by Eqs. (\ref{eq:relat_dens_reg},\ref{eq:YoungIso}), we find an expression of $k_m$ in terms of the acceleration $g$, the density $\rho_{up}$, and the properties of the material the foam is made of,
\begin{equation}
k_m = \frac{\sqrt{3}}{16} g \frac{   \rho_{up}-\frac{2}{\sqrt{3}}\frac{t}{l} \rho_s   }{E_s} \left( \frac{l}{t} \right)^3.
\end{equation}

The present results have been derived for a simplified model of foam. Yet, all foams seem to exhibit an elastic phase at the beginning of their strain-stress curve, be it in 2D \cite{Gibson1982_2D} or even in 3D \cite{Gibson1982_3D}. Since our key result, namely the reduction of the RTI growth rate in the elastic phase, relies on the existence of such a phase, our conclusion is likely valid for most foams. Only the Young modulus $E$  involved in Figure \ref{fig:streerstrain} and Section \ref{sec:elastic} needs to be adapted to the specific foam under scrutiny.

As explained in the introduction, this work aimed at filling the ``intact'' end of the gap between  intact and homogenised foam. In the context of ICF, the foam is basically a plasma, not a solid, after the laser beam propagated through it. So the three phases discussed here, namely, elastic, plastic and fracture, may likely be irrelevant.  Consequently, the RTI behaviour in a non-uniform foam plasma would be highly interesting to explore.

\section{Acknowledgments}
This work was supported in part by the U.S. Department of Energy NNSA MIT Center-of-Excellence under Contract DE-NA0003868. A.B. acknowledges support by the Spanish Ministerio de Ciencia, Innovaci\'{o}n y Universidades Grant No. PID2024-157933OA-I00 and by EUROfusion Grant No. CfP-FSD-AWP24-ENR-03-CEA-03. A.B. Thanks the MIT PSFC for its hospitality in June 2025.


\bibliography{BibBret}

\end{document}